\documentclass[letterpaper, 10 pt, conference]{ieeeconf}  

\IEEEoverridecommandlockouts                              

\usepackage{multirow}
\usepackage{graphicx}
\usepackage{caption}
\usepackage{url}

\title{\LARGE \bf
Work Smarter Not Harder: Simple Imitation Learning with CS-PIBT Outperforms Large Scale Imitation Learning for MAPF
}

\author{Rishi Veerapaneni$^{*1}$, Arthur Jakobsson$^{*1}$, Kevin Ren$^{1}$, Samuel Kim$^{2}$, Jiaoyang Li$^{1}$, Maxim Likhachev$^{1}$
\thanks{*Equal Authorship}
\thanks{$^{1}$ Carnegie Mellon University, USA. {\small \tt \{{vrishi, ajakobss, kevinren, maxim, jiaoyangli\} @cmu.edu}}}
\thanks{$^{2}$ Solon High School, USA. {\small \tt \{{samuelkim0617@gmail.com\}}}}
}

\begin{document}

\maketitle
\thispagestyle{empty}
\pagestyle{empty}

\begin{abstract}

Multi-Agent Path Finding (MAPF) is the problem of effectively finding efficient collision-free paths for a group of agents in a shared workspace. The MAPF community has largely focused on developing high-performance heuristic search methods. Recently, several works have applied various machine learning (ML) techniques to solve MAPF, usually involving sophisticated architectures, reinforcement learning techniques, and set-ups, but none using large amounts of high-quality supervised data.
Our initial objective in this work was to show how simple large scale imitation learning of high-quality heuristic search methods can lead to state-of-the-art ML MAPF performance. However, we find that, at least with our model architecture, simple large scale (700k examples with hundreds of agents per example) imitation learning does \textit{not} produce impressive results. Instead, we find that by using prior work that post-processes MAPF model predictions to resolve 1-step collisions (CS-PIBT), we can train a simple ML MAPF model in minutes that dramatically outperforms existing ML MAPF policies.
\textit{This has serious implications for all future ML MAPF policies (with local communication) which currently struggle to scale.} 
In particular, this finding implies that future learnt policies should (1) always use smart 1-step collision shields (e.g. CS-PIBT), (2) always include the collision shield with greedy actions as a baseline (e.g. PIBT) and (3) motivates future models to focus on longer horizon / more complex planning as 1-step collisions can be efficiently resolved. 
\end{abstract}

\section{INTRODUCTION}
The rise of cheaper and more capable individual robots has increased the prevalence of multi-robot systems. One core problem in multi-robot systems is Multi-Agent Path Finding (MAPF) which requires finding collision free paths for the agents in a shared workspace. Without careful consideration, solution paths can be extremely inefficient or take an impractical amount of time to compute.

The MAPF community over the past decade has developed a variety of high-performing suboptimal heuristic search methods that can scale to hundreds of agents in gridworld \cite{li2021eecbs,li2022mapf-lns2,okumura2024lacam3}. However, the majority of these methods require a centralized planner and planning time on the order of seconds (if not tens of seconds) to find a solution.
Additionally, one potential limitation of centralized planners is that they require global communication between agents, which can be impractical.

An appealing alternative approach to using heuristic search methods is to use machine learning to directly learn a MAPF policy for each agent. Depending on implementation, this could theoretically be decentralized (e.g. each agent running its own policy and communicating with a small subset of neighboring agents), fast (just the time required for a neural network inference), and result in efficient paths. There have been a variety of ML works that have proposed different MAPF policies and showed faster runtime and scalability against \textit{optimal} search methods and prior ML works.

The vast majority of these works have used reinforcement learning (RL) to learn decentralized policies for agents \cite{sartoretti2019primal,ma2021dhc,ma2022dcc,lin2023sacha,skrynnik2024learn-to-follow,tang2024eph}. These methods have tried a variety of reward functions, training curriculums, model inputs, complex architectures, etc. However, for regular (single-shot) MAPF, these methods trained with a fairly small number of agents (e.g. 16) or agent density (e.g. 2\%). Thus, when faced with larger agent densities (e.g. $>$10\%), current RL MAPF methods fail and perform substantially worse than existing \textit{suboptimal} heuristic search solvers.

\begin{table*}[t!]
\vspace{1em}
\centering
\resizebox{0.95\textwidth}{!}{
\begin{tabular}{|c||cccccccc|c|}
\hline
Prior Work & PRIMAL & GNN & DHC & DCC & MAGAT & SACHA & SCRIMP & EPH & Ours \\ \hline
Training Maps & Random & Random & Random & Random & Random & Random & Random & Random & Moving AI \\ \hline
Testing Maps & Random & Random & Random & Random & Random & \begin{tabular}[c]{@{}c@{}} Random, den312d \\ warehouse \end{tabular} & Random & \begin{tabular}[c]{@{}c@{}} Random, den312d \\ warehouse \end{tabular} & Moving AI \\ \hline
Max Density & 9\% & 3\% & 6\% & 6\% & 4\% & 7\% & 11\% & 6\% & \textgreater{}40\% \\ \hline
\end{tabular}}
\caption{We list prior ML MAPF works' training maps, testing maps, and maximum agent density evaluated with non-trivial success rates. We limit this table to standard MAPF and not other variants (e.g. Lifelong-MAPF). We train and evaluate on nearly the full MAPF Moving AI benchmark, which contains a diverse set of maps and is used by heuristic search methods \cite{stern2019mapfbenchmark} as opposed to prior work which focus on random maps. We can also handle higher agent density situations.}
\label{tab:prior-work}
\vspace{-1em}
\end{table*}

Our main objective in this paper is to investigate the performance of ML models trained via simple and scalable imitation learning by constructing a large training dataset (700k+ examples with hundreds of agents per example) using a strong heuristic search solver. Few prior ML MAPF works have used supervised learning, and those that did used weak solvers \cite{li2020gnn,li2021magat}. 
However, we found that our trained model performed fairly poorly compared to existing heuristic search solvers. 
Instead, we found that when using prior work on post-processing model outputs with a 1-step collision resolution technique (CS-PIBT) \cite{veerapaneni2024improving_mapf_policies_with_search}, we were able to train a ``state-of-the-art" ML MAPF model with 100x less data and in less than 5 minutes of training time. 
We additionally conducted several studies on how the quality and quantity of the supervised learning datasets effected performance.

Thus, our main finding is that future ML MAPF methods can benefit from ``working smarter and not harder." 
We note that current methods have been hampered by 1-step collisions \cite{wang2023scrimp,tang2024eph}. \cite{veerapaneni2024improving_mapf_policies_with_search} demonstrated how CS-PIBT is a model agnostic way to improve model performance and touches upon how this could ease the learning problem. Our results empirically and emphatically demonstrate this point.

Our paper thus has several important findings for future work.
First, and shared with \cite{veerapaneni2024improving_mapf_policies_with_search}, \textbf{future ML MAPF policies should always use a smart collision shield}, e.g. CS-PIBT (or future ones that get developed). Second, \textbf{all future works should be required to compare to the smart collision shield with greedy actions, i.e. PIBT,} as this is a trivial solution the model could learn.
Third, since using CS-PIBT means that 1-step collisions are resolved, \textbf{future ML MAPF models can focus on longer-horizon reasoning}. 
In particular, we qualitatively observe that target conflicts are a key longer-horizon problem that negatively affected performance. 
Last, \textbf{large scale imitation learning is fast and feasible given existing strong heuristic search solvers}. Although our results showed that it did \textit{not} help performance when using CS-PIBT, it is likely that future work using more sophisticated model architectures and curated datasets can learn effective longer-horizon behavior from large scale imitation learning. Concurrent work MAPF-GPT \cite{andreychuk2024mapf-gpt} also uses large scale imitation learning to train a (transformer) MAPF policy. 


\section{BACKGROUND}
\subsection{Problem Formulation}
Multi-Agent Path Finding (MAPF) is the problem of finding collision-free paths for a group of $N$ agents ${i = 1, \dots, N}$, that takes each agent from its start location $s_i^{start}$ to its goal location $s_i^{goal}$. In traditional 2D MAPF, the environment is discretized into grid cells, and time is broken down into discrete timesteps. Agents are allowed to move in any cardinal direction or wait in the same cell. A valid solution is a set of agent paths $\Pi = \{ \pi_1, ..., \pi_N \}$ where $\pi_i[0] = s_i^{start}$, $\pi_i[T_i] = s_i^{goal}$ where $T_i$ is the maximum timestep of the path for agent $i$.
Critically, agents must avoid vertex collisions (when $\pi_i[t] = \pi_{j \neq i}[t]$) and edge collisions (when $\pi_i[t] = \pi_j[t+1] \wedge \pi_i[t+1]=\pi_j[t]$) for all timesteps $t$. The typical objective in MAPF is to find a solution $\Pi$ that minimizes $\sum_{i=1}^N |\pi_i| = \sum_{i=1}^N \sum_{t=0}^{T_i-1} c(s_i^t,s_i^{t+1})$.

\subsection{Related Work}
\subsubsection{Heuristic Search MAPF Solvers}
There are a variety of high-performance suboptimal, centralized, heuristic search MAPF methods. ECBS \cite{barer2014suboptimalecbs}, EECBS \cite{li2021eecbs}, W-EECBS \cite{effectiveCBS} are suboptimal variants of optimal Conflict-Based Search (CBS) \cite{sharon2015cbs} and can solve 2D MAPF instances containing hundreds of agents. 
MAPF-LNS \cite{li2021anytime_lns} and MAPF-LNS2 \cite{li2022mapf-lns2} find an initial suboptimal solution fast and then uses local neighborhood search using ``destroy" and ``repair" operators to iteratively improve solutions.
The LaCAM / LaCAM* methods \cite{okumura2023lacam,okumura2023lacam2,okumura2024lacam3} perform a joint-space search and use PIBT \cite{okumara2022pibt_journal} and lazy successor generation to quickly find MAPF solutions, scaling up to thousands of agents.

All of these methods, except for PIBT, are centralized and require a single solver to plan paths for all agents. This imposes strict communication requirements which can be a bottleneck in real-world systems (e.g. warehouses) \cite{communicationReview2022}.

\subsubsection{Machine Learning MAPF Solvers}
Machine Learning MAPF Solvers can be broadly divided into two categories. First, there are methods that incorporate machine learning as a component of existing heuristic search methods. These methods can retain the benefits of heuristic search methods (e.g. completeness or bounded suboptimality) while speeding them up. \cite{yu2023learnConflictHeuristic} learns a heuristic function to accelerate ECBS while \cite{yan2024neural-lns} speeds up MAPF-LNS training a network to intelligently determine the destroy operator.

Second, there are methods that directly learn a single-step MAPF policy. These methods have no guarantees, but can potentially be decentralized and faster than heuristic search methods. There are a large number of works that train such policies across standard MAPF as well as other variants (e.g. Lifelong-MAPF where agents get new goals when they reach their current goal).
In rough chronological order, DHC \cite{ma2021dhc}, DCC \cite{ma2022dcc}, SACHA \cite{lin2023sacha}, Follower \cite{skrynnik2024learn-to-follow}, and EPH \cite{tang2024eph} all solely use reinforcement learning to train individual agent policies.
Graph Neural Networks for Decentralized Multi-Robot Path Planning \cite{li2020gnn} and MAGAT \cite{li2021magat} solely use imitation learning to train a Graph Neural Network \cite{graph_nn_2017} policy. Here, agents are nodes in the graph that have ``edges" with close-by agents and can communicate / message-pass with such agents.
PRIMAL \cite{sartoretti2019primal}, PRIMAL2 \cite{damani2021primal2}, PICO \cite{li2022pico}, and SCRIMP \cite{wang2023scrimp} utilize a combination of reinforcement learning and imitation learning.

These approaches explore a variety of different input features, neural architectures, and inter-agent communication mechanisms. They show improvements (better success rates, scalability, runtime) over their respective prior ML MAPF methods and certain heuristic search methods. 

However, existing work has key limitations. First, they mostly compare against weak (e.g. OrDM* \cite{ferner2013ordm}) or optimal (e.g. CBS) heuristic search solvers. Modern suboptimal heuristic search methods would significantly beat these methods.
Second, prior methods on standard (e.g. not Lifelong) MAPF primarily train and test on maps with randomly sampled obstacles (Table \ref{tab:prior-work}).
On the other hand, heuristic search methods typically evaluate on the MAPF Moving AI Benchmark which contains a diverse set of scenarios (den312d and warehouse in Table \ref{tab:prior-work} are two of these maps) \cite{stern2019mapfbenchmark}.
Third, these methods train and test with limited agent density. To our knowledge, the maximum agent density with a non-trivial success rate is approximately 10\%.

Our work attempts to improve all three of these components by comparing against modern heuristic search methods, training (and testing) on a variety of maps using large scale imitation learning, and getting non-trivial success rates at agent densities of over 40\%

MAPF-GPT \cite{andreychuk2024mapf-gpt}, a concurrent work to our research, utilizes a similar large-scale imitation learning approach and shows how this outperforms existing ML MAPF works.
Our work shows that, given a smart collision shield, we can outperform existing ML MAPF performance without large scale data.

\begin{figure*}[t]
    \centering
    \includegraphics[width=0.9\linewidth]{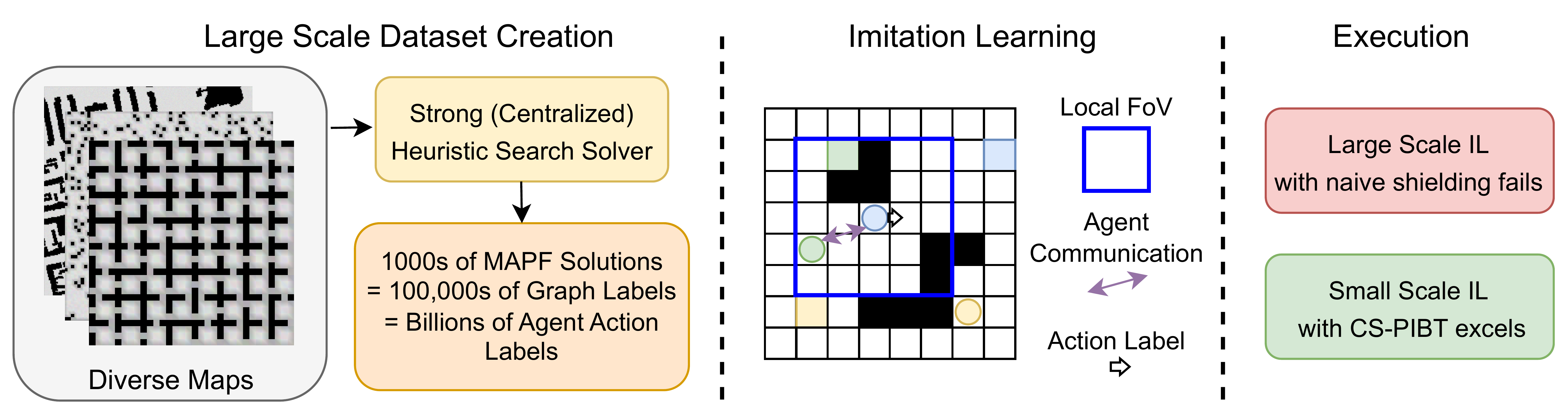}
    \caption{Left: From a diverse sets of map, we can use existing strong (centralized) heuristic search solvers to solve instances with hundreds of agents. This can easily lead to thousands of MAPF solutions. Each solution contains a sequence of timesteps, where each timestep can be viewed as a ``graph" of agents where each agent has a label. Middle: We can then train a local decentralized policy that takes as input a local field of view (FoV) and tries to predict the action label. Additionally, the agent can communicate to agents nearby within the FoV. Right: With naive collision shielding, our performance is not impressive. With CS-PIBT, we find that we do not need large data and a small scale imitation learning is able to get good performance.}
    \label{fig:imitation-learning}
    \vspace{-1em}
\end{figure*}

\subsubsection{Collision Shielding during Execution}
When executing a model during test-time execution, it is possible for agents to propose actions that collide with other agents. Thus, actions must be post-processed with a ``\underline{C}ollision-\underline{S}hield" to prevent collisions during execution \cite{veerapaneni2024improving_mapf_policies_with_search,li2020gnn}. The majority of prior work have used naive collision shields which freeze agents that propose colliding actions. 
SCRIMP and EPH improve upon this by having ``lower" priority agents try to resample actions when colliding with ``higher" priority agents.
However, this approach becomes ineffective when several agents are colliding with each other. 

Prior work proposes CS-PIBT, which post-processes the network's probabilities and uses PIBT as a collision shield to resolve 1-step collisions between agents or obstacles, and found that this boosts the performance of learnt ML MAPF policies \cite{veerapaneni2024improving_mapf_policies_with_search}. From a high-level, CS-PIBT/PIBT uses agent priorities with priority inheritance and backtracking to resolve collisions. In particular, if a high priority agent moves into a low-priority agent, the low-priority agent ``inherits" the high priority and moves later, with priority cascading and allowing several agents in congestion to move. Since PIBT is decentralized as it only requires agents in collision to communicate, using CS-PIBT retains the decentralized property of the model. Our experiments find that using CS-PIBT significantly changes the learning problem.

\subsubsection{Large-Scale Learning in Machine Learning}
Our work is inspired by the success of large-scale learning in the broader machine learning field. In particular, the success of the vision and language community has shown how training on large datasets can lead to state-of-the-art performance on a variety of tasks \cite{lin2024coco_dataset,deng2009imagenet,radford2021clip_model,kirillov2023segment_anything,zhu2015bookcorpus,radford2019gpt2_webtext,raffel2019colossalcrawledcorpus,brown2020gpt3}. 


\section{Method}
Our objective is to learn a decentralized MAPF policy that can work well on a variety of maps.
Our main idea is that we can utilize existing strong centralized heuristic search methods to gather training data, and then use imitation learning to train the policy.

\subsection{Collecting Training Data}
A gridworld MAPF scenario consists of a map with binary obstacles, and a set of agents, each with a unique start and goal location. To obtain supervised data (i.e. collision free paths for all agents), we can use any standard MAPF heuristic search method. Since we are training a single-step policy, we can use every timestep on the solution path as an input scenario (i.e. map, agents with start and goal locations) and label it with the next action in the solution path. 

We chose to use EECBS \cite{li2021eecbs} to create the supervised data as EECBS with SIPPS \cite{sipp2011,li2022mapf-lns2} is the current strongest bounded suboptimal MAPF solver. The ability to control the bounded suboptimality allows us to explicitly control the solution quality of collected data. 


Our emphasis is to train and test on a diverse set of scenarios. To that end, we use the MAPF Moving AI dataset \cite{stern2019mapfbenchmark} which contains a diverse set of empty, random obstacles, game, city, and warehouse maps.

\subsection{Decentralized Single-Step Local ML MAPF Policy}
We follow standard ML MAPF approaches and predict a single-step action probability for each agent given limited inter-agent communication.
To encourage generalizing to maps of different sizes as well as reduce the complexity of the learning problem, we follow the standard ML MAPF approach and limit the information the agent receives to a field of view (FoV) centered at the agent. 

To have a decentralized policy, we again follow prior ML MAPF approaches of using a Graph Neural Network (GNN) \cite{graph_nn_2017,li2020gnn,li2021magat}.
In this GNN structure, ``nodes" are agents with ``edges" connecting agents. Each agent outputs a probability distribution over 5 actions corresponding to the 4 cardinal directions plus a wait action.
We enforce the decentralized policy of the agent by only letting it communicate with the $M$ nearest agents within the agent's FoV, i.e. an agent ``node" will have at most $M$ ``edges." Thus, if all agents are outside of an agent's FOV, the agent will have no edges and act independently.
Under this setup, ``edges" explicitly denote communication between agents as the GNN policy requires message passing across ``nodes." We use $M=5$.

\subsubsection{Model Architecture}
We note our goal is \textit{not} to propose a specific model architecture. We used a standard SageConv GNN \cite{hamilton2017sagegraphconv} from Pytorch Geometric \cite{pytorchgeometric}. We did some small hyper-parameter tuning of our architecture (e.g. activation, number of layers, layer size, layer norm, dropout) but found that performance did not change significantly given a sufficiently powerful GNN (e.g. SageConv as opposed to other weaker GNN architectures we tried). 

Our GNN first takes in node (agent) features within a FoV. We did not utilize edge features.
We define the FoV to have radius $R$, leading to an FoV of size $D*D$, where $D=2R+1$, centered at the agent's current position $(x_c,y_c)$. We pass in three $D*D$ images centered at the agent's current position. The first image is a binary image that encodes obstacles (i.e. 1 = obstacle, 0 = free space). The second image $I_h$ is a normalized heuristic heatmap of the Backward Dijkstra heuristic $h$, where $I_{h}[x,y] = \frac{h(x,y)-h(x_c,y_c)}{2R}$. We additionally clamp $I_h$ to contain values between -1 and 1. Our third $D*D$ image was a binary agent occupancy map with a 1 if an agent is at that current location. We additionally found it useful to pass in a binary vector of size 5 corresponding to the 5 actions with a 1 if that action minimizes the agent's heuristic.

We encode the $(3,D,D)$ image via a CNN, flatten the embedding, and concatenate it with the size 5 binary vector. We pass the embedding into two different linear layers to get a node and message embedding.
After that we used three SageConv graph convolution layers (with Layer Norms between layers) for more message passing between agents. We finally pass the embedding through two linear layers to obtain the action probability distribution.


\subsection{Training}
The model minimizes standard cross-entropy loss given the action labels from the supervisor heuristic search method. Note we do not apply CS-PIBT in training as this would significantly slow down the training process.


\section{EXPERIMENTS}
\subsection{Training and Testing Set-Up} \label{sec:dataset}
We consider 27 of the 31 MAPF Moving AI benchmark maps \cite{stern2019mapfbenchmark} in our experiments. 
We omitted four maps (brc202d, orz900, maze-128-128-1, maze-128-128-10) as they had issues in our data collection pipeline (e.g. very large or EECBS had issues solving).

We divide the benchmark into 19 training maps and 8 held-out testing maps. We also added 6 random obstacle maps of different sizes with 10\% obstacle occupancy and 1 small corridor map to the training set. We collect training data by running EECBS on the training maps with a suboptimality of $2$ and a timeout of $2$ minutes, on 20 to 1000 agents depending on the map size. 
We experiment with collecting 1, 4, 16, and 128 randomly generated start-goal scenarios per (map, num agents).
Since each timestep of a MAPF solution can be used as an individual training example (since we are only doing one-step predictions), our maximum dataset size (with 128 scenarios per (map, num agents)) contains 712,751 full graph training examples. 
We note that each graph training example contains the positions and labels of all agents throughout the entire map at a given timestep. Thus, each graph contains data for many agents, which greatly expands the amount of agent-level information in our training set.
Since each MAPF instance is independent, we can collect data in parallel. By running 64 EECBS solvers in parallel, it took approximately 
90 minutes to collect the 128-scene dataset.

We use $R=4$ (resulting in a FoV of $9*9$) as we found that larger values reduce performance.
Training is very fast. On the entire 128-scene dataset, training for two epochs takes approximately 30 minutes with a batchsize of 64 on a v100-32GB GPU (taking up 1GB GPU Memory). Training on the 1-scene dataset took 4 minutes. Training for two epochs was sufficient to reach a test accuracy of about 89\% in all instances. Training for ten epochs did not meaningfully change results.
For comparison, EPH takes 20 hours on an RTX A6000 GPU, and SCRIMP requires 8 hours on 4 3090 GPUs in order to converge. 

We evaluate our method with a 2 minute timeout or 3x makespan limit on the 27 selected MAPF Moving AI benchmark maps, and their provided 25 scenes per map, with agent increments of 100. Thus, our evaluation has over 3,000 scenarios not seen in training. 

\begin{figure*}[t!]
    \centering
    \includegraphics[width=0.99\linewidth]{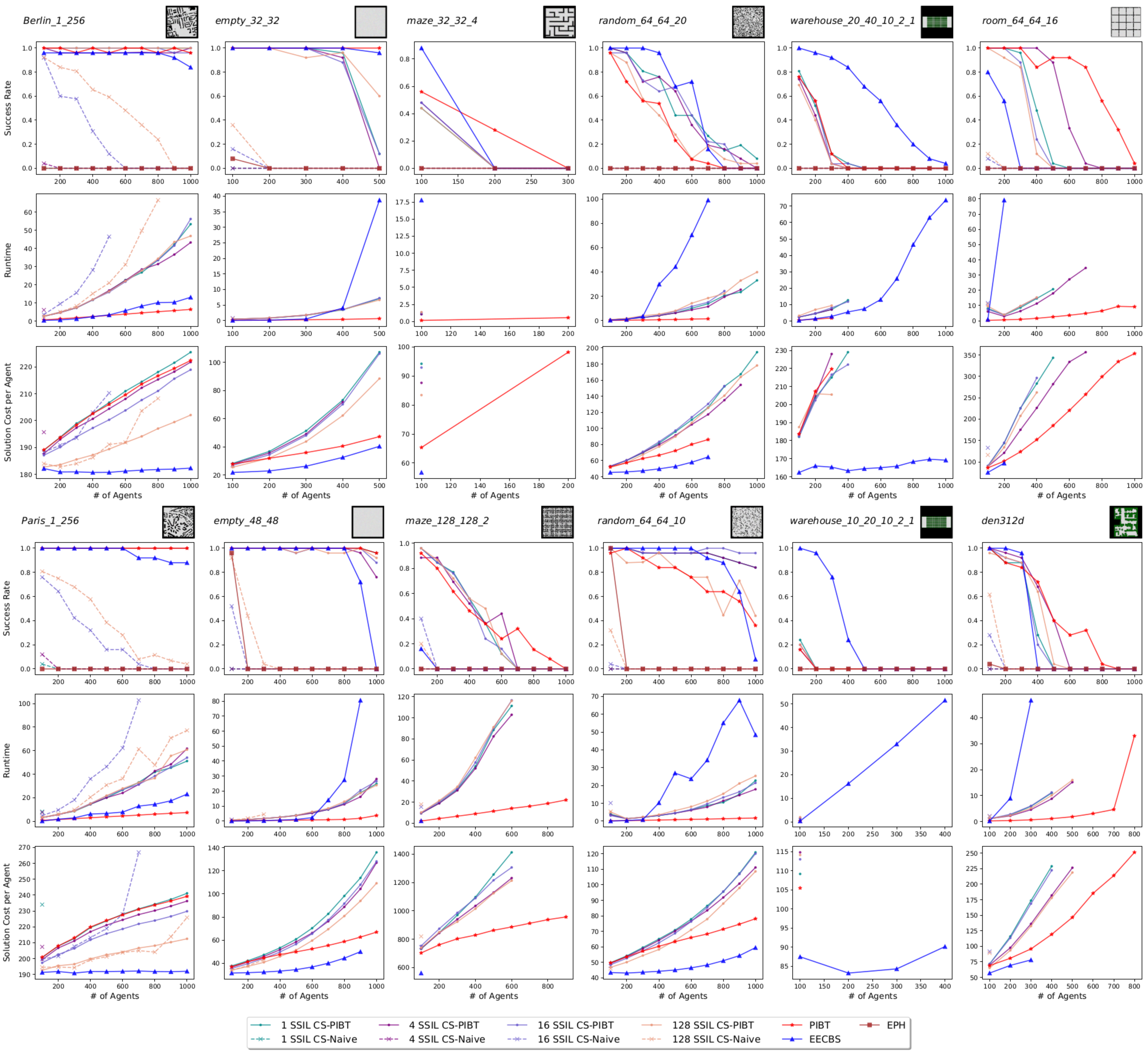}
    \vspace{-0.5em}
    \caption{We plot the success rate, per-agent solution cost, and runtime for each method. The first row maps are seen during training, but evaluated with new start-goal positions, while the bottom maps are unseen maps. We see how our SSIL with CS-PIBT significantly outperforms EPH. It is also on-par with PIBT and EECBS with performance dependent on the map.}
    \label{fig:main-results}
    \vspace{-1em}
\end{figure*}

\subsubsection{Baselines}
In terms of competing ML methods, we compare against EPH \cite{tang2024eph}, which is a recent state-of-the-art RL MAPF method that was shown to outperform prior RL MAPF methods \cite{wang2023scrimp,lin2023sacha,ma2022dcc,ma2021dhc,sartoretti2019primal}. We evaluate EPH's pretrained model (trained on random maps) with a 5 minute timeout across all maps. We additionally plot the results of EECBS \cite{li2021eecbs}, which is a state-of-the-art bounded suboptimal (centralized) heuristic search MAPF solver, with a 2 minute timeout and all optimizations. Note that LaCAM* was shown to have a perfect success rate under 1-minute across all our tested maps and we therefore do not include it as it dominates all methods \cite{okumura2023lacam2,okumura2024lacam3}.

A primary motivation for learning MAPF policies is that they can be decentralized compared to centralized heuristic search methods. However, all prior works published after PIBT (first version in 2019) failed to notice that PIBT is a theoretically decentralized method \cite{okumara2022pibt_journal}. PIBT only requires communication between conflicting agents and plans only 1-step actions. Thus, PIBT with a 2 minute timeout or 3x makespan limit is our main heuristic search competitor.

\subsection{Results} \label{sec:results}
Figure \ref{fig:main-results} shows results for our models (abbreviated as SSIL for \underline{S}imple and \underline{S}calable \underline{I}mitation \underline{L}earning) and baseline methods. The top row contains maps seen during training, but are evaluated with unseen start-goal locations. The bottom row tests the model's ability to generalize to new maps.

\subsubsection{Comparison with Baselines}
We first contrast our models' performance against EPH. SSIL with CS-PIBT (solid, several colors) is able to significantly outperform EPH (solid, maroon), which is unable to solve the minimum number of agents in most cases. 
Although part of this could be attributed to testing EPH on more diverse maps than it was trained on (since EPH's pre-trained model was only trained on random maps), EPH also scales poorly on the random maps.

Second, we compare against EECBS (solid blue). Since EECBS was the centralized heuristic search method used to provide labels, instances where SSIL scales larger than EECBS (e.g. empty-48-48) showcases how the learnt model can even outperform its supervising method. 
We delay our comparison to PIBT to the next section.


\subsubsection{Working Smarter and Not Harder with CS-PIBT}
First, we focus on the performance of SSIL with naive collision shielding (dashed lines, SSIL CS-Naive). As we increase the amount of training data, e.g. going from 1-SSIL to 16 to 128, we see that the performance of SSIL improves on sparser maps (e.g. Berlin\_1\_256) or with fewer agents (e.g. den312d with 100). In general, though, SSIL with CS-Naive is unable to perform well across the majority of instances we tested. 

However, when we replace CS-Naive with CS-PIBT (solid lines), we see that the story significantly changes. Large data is \textit{not} required and we can train an unquestionably ``state-of-the-art" ML MAPF model with 1-scene's worth of data which only required 4 minutes of training time. We emphasize we run the exact same trained models, but just resolve 1-step collisions using CS-PIBT instead of CS-Naive. 

This leads to two observations, one positive and one potentially cynical.
Our positive observation is that this implies that one main bottleneck in learning MAPF policies is resolving 1-step collisions. 
The standard large data solution of increasing training data and model complexity will likely take significant time and compute.
Using CS-PIBT enables more data-efficient approaches.

Our potentially cynical observation is that it is trivially easy to obtain a current ``state-of-the-art" ML MAPF model using CS-PIBT. Any model that simply prefers the greedy action that minimizes the heuristic would exactly mimic PIBT. This is easy given that standard inputs into ML MAPF models contain the backward Dijkstra heuristic.
As seen in Figure \ref{fig:main-results}, PIBT completely dominates existing ML MAPF works. Additionally, our ``state-of-the-art" performance is strongly correlated with PIBT's performance. This implies our model may not be successfully learning long-horizon coordination but rather shorter / greedier behavior. Thus, future work should always use intelligent collision shields (e.g. CS-PIBT) and also compare against the collision shield with the greedy heuristic (e.g. PIBT) to accurately gauge the model's relative contribution.




\begin{figure}[t]
    \centering
    \includegraphics[width=0.95\linewidth]{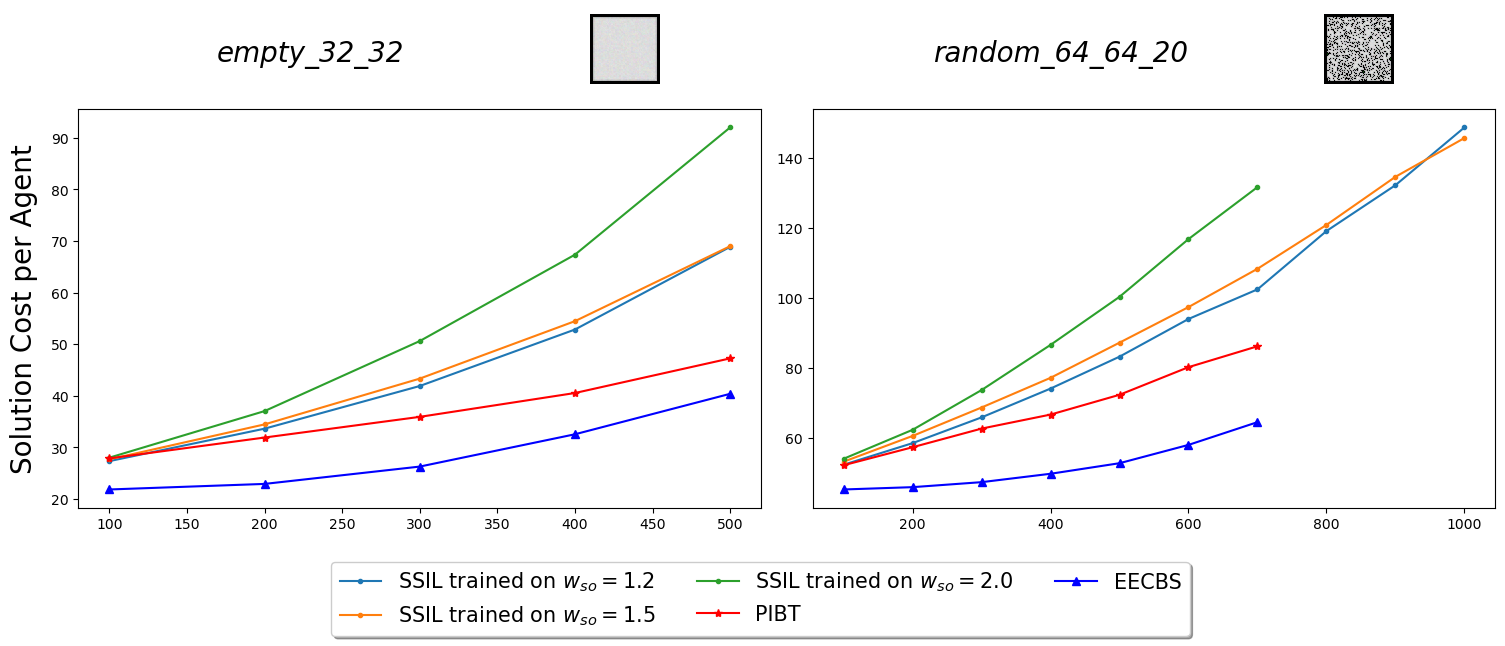}
    \caption{We see how SSIL produces solutions with better solution quality when trained on data collected from EECBS run with lower suboptimality.}
    \label{fig:suboptimality-results}
    \vspace{-1.5em}
\end{figure}

\subsubsection{Impact of Data Quality on Performance}
Since EECBS is a bounded suboptimal algorithm, we can control the quality of the data by changing its suboptimality factor $w_{so}$ used during data collection. We experimented with training SSIL on data collected from EECBS with $w_{so}=1.2, 1.5, 2$. Figure \ref{fig:suboptimality-results} shows that training on a tighter suboptimality does indeed result in a model that produces better solution quality.
Although not plotted, we saw that the success rates between the different methods did not differ systematically between SSIL trained on different suboptimalities.



\subsection{Observations and Recommendations}
\subsubsection{Failure Modes}
We qualitatively observed two main modes of failure when using CS-PIBT. First, the most common failure we observed were situations where an agent was resting at its goal location and blocking another agent from passing through. These are typically referred to as ``Target Conflicts" in MAPF heuristic search literature \cite{srli2021}. The stuck agents would frequently jitter back-and-forth when in these situations. Second, in certain maps with bottlenecks, agents would get stuck in congestion with many other agents.

We specifically tried to tackle Target Conflicts but found it difficult to address \cite{wang2023scrimp}.
We tried some rounds of DAgger \cite{ross2011dagger} but found that it did not significantly boost performance.
One interesting interpretation of these target conflicts is that the issue is not that the agent is resting at its goal, but rather that it entered its goal location too early. EECBS, with its global reasoning, could have that agent wait at a non-blocking location to let the other agent enter the goal first. 
We tried inputting information about other agents' goal locations (via a goal image) but found it did not improve model performance. 

To address the bottleneck issue, newer work in heuristic search has proposed using non-uniform edge costs or guidance \cite{chen2024trafficflow,zhang2024ggo}. Future ML MAPF works should exploit this literature and be amenable to such works \cite{skrynnik2024learn-to-follow}.


\subsubsection{Longer-Horizon Reasoning}
We note that both previously mentioned issues are \textit{longer} horizon reasoning problems as CS-PIBT eliminates the need to focus on 1-step collision avoidance.
Additionally, one surprising result we saw was that SSIL had a worse solution quality than PIBT, even though it was trained on EECBS (which can be seen to have clearly better costs than PIBT in Fig \ref{fig:main-results}). PIBT is extremely greedy, while EECBS handles long term dependencies. 
This implies that learning policies on \textit{local} information might be unable to allow reasoning about longer horizons that are vital for EECBS's performance. 
Future work should investigate mechanisms to enable longer-horizon reasoning.

\subsubsection{Limitations of our Model} \label{sec:limitations}
One potential key limitation of our model is that it decomposed the MAPF problem into independent decisions per timestep (e.g. Markovian decisions). This is clearly not true in MAPF and could contribute to part of the poor performance we observed (e.g. jittering at target conflicts). However, several prior works have used this assumption \cite{li2020gnn,li2021magat,tang2024eph} while other have included sophisticated recurrent models \cite{damani2021primal2,wang2023scrimp} and all currently struggle to scale. Future methods tackling longer-horizon reasoning will likely benefit from recurrent models.

\section{CONCLUSIONS}
This work investigates simple and scalable imitation learning for MAPF.
We find that simple large-scale imitation learning improvements still has significant problems resolving 1-step collisions (at least using our model architecture and inputs). However by combining it with prior work CS-PIBT \cite{veerapaneni2024improving_mapf_policies_with_search}, which resolves 1-step collisions, we can get ``state-of-the-art" ML-MAPF performance across a wide variety of maps in minutes via imitation learning. 

This finding, along with our quantitative and qualitative observations, has four important implications for future work. First, we advise future work to use CS-PIBT (or other smart collision shields). Second, since a trivial solution of using a model with CS-PIBT is learning greedy behaviors, always compare against PIBT. Third, ML models can attempt to learn longer-horizon behaviors, as CS-PIBT resolves local collisions. Lastly, scalable imitation learning, with the existence of strong heuristic search methods, is easy and a promising way to train future ML MAPF models.





\bibliographystyle{plain}
\bibliography{ref}

\begin{thebibliography}{10}

\bibitem{andreychuk2024mapf-gpt}
Anton Andreychuk, Konstantin Yakovlev, Aleksandr Panov, and Alexey Skrynnik.
\newblock Mapf-gpt: Imitation learning for multi-agent pathfinding at scale.
\newblock {\em arXiv preprint arXiv:2409.00134}, 2024.

\bibitem{barer2014suboptimalecbs}
Max Barer, Guni Sharon, Roni Stern, and Ariel Felner.
\newblock Suboptimal variants of the conflict-based search algorithm for the multi-agent pathfinding problem.
\newblock In {\em Seventh Annual Symposium on Combinatorial Search}, 2014.

\bibitem{brown2020gpt3}
Tom~B. Brown, Benjamin Mann, Nick Ryder, Melanie Subbiah, Jared Kaplan, Prafulla Dhariwal, Arvind Neelakantan, Pranav Shyam, Girish Sastry, Amanda Askell, Sandhini Agarwal, Ariel Herbert{-}Voss, Gretchen Krueger, Tom Henighan, Rewon Child, Aditya Ramesh, Daniel~M. Ziegler, Jeffrey Wu, Clemens Winter, Christopher Hesse, Mark Chen, Eric Sigler, Mateusz Litwin, Scott Gray, Benjamin Chess, Jack Clark, Christopher Berner, Sam McCandlish, Alec Radford, Ilya Sutskever, and Dario Amodei.
\newblock Language models are few-shot learners.
\newblock {\em CoRR}, abs/2005.14165, 2020.

\bibitem{chen2024trafficflow}
Zhe Chen, Daniel Harabor, Jiaoyang Li, and Peter Stuckey.
\newblock Traffic flow optimisation for lifelong multi-agent path finding.
\newblock In {\em Proceedings of the AAAI Conference on Artificial Intelligence (AAAI)}, pages 20674--20682, 2024.

\bibitem{damani2021primal2}
Mehul Damani, Zhiyao Luo, Emerson Wenzel, and Guillaume Sartoretti.
\newblock Primal $ \_2 $: Pathfinding via reinforcement and imitation multi-agent learning-lifelong.
\newblock {\em IEEE Robotics and Automation Letters}, 6(2):2666--2673, 2021.

\bibitem{deng2009imagenet}
Jia Deng, Wei Dong, Richard Socher, Li-Jia Li, Kai Li, and Li~Fei-Fei.
\newblock Imagenet: A large-scale hierarchical image database.
\newblock In {\em 2009 IEEE Conference on Computer Vision and Pattern Recognition}, pages 248--255, 2009.

\bibitem{ferner2013ordm}
Cornelia Ferner, Glenn Wagner, and Howie Choset.
\newblock Odrm* optimal multirobot path planning in low dimensional search spaces.
\newblock In {\em 2013 IEEE International Conference on Robotics and Automation}, pages 3854--3859, 2013.

\bibitem{pytorchgeometric}
Matthias Fey and Jan~E. Lenssen.
\newblock Fast graph representation learning with {PyTorch Geometric}.
\newblock In {\em ICLR Workshop on Representation Learning on Graphs and Manifolds}, 2019.

\bibitem{graph_nn_2017}
Fernando Gama, Antonio~G. Marques, Geert Leus, and Alejandro Ribeiro.
\newblock Convolutional neural network architectures for signals supported on graphs.
\newblock {\em IEEE Transactions on Signal Processing}, 67(4):1034–1049, February 2019.

\bibitem{communicationReview2022}
Jennifer Gielis, Ajay Shankar, and Amanda Prorok.
\newblock A critical review of communications in multi-robot systems.
\newblock {\em CoRR}, abs/2206.09484, 2022.

\bibitem{hamilton2017sagegraphconv}
William~L. Hamilton, Zhitao Ying, and Jure Leskovec.
\newblock Inductive representation learning on large graphs.
\newblock In Isabelle Guyon, Ulrike von Luxburg, Samy Bengio, Hanna~M. Wallach, Rob Fergus, S.~V.~N. Vishwanathan, and Roman Garnett, editors, {\em Advances in Neural Information Processing Systems 30: Annual Conference on Neural Information Processing Systems 2017, December 4-9, 2017, Long Beach, CA, {USA}}, pages 1024--1034, 2017.

\bibitem{kirillov2023segment_anything}
Alexander Kirillov, Eric Mintun, Nikhila Ravi, Hanzi Mao, Chlo{\'{e}} Rolland, Laura Gustafson, Tete Xiao, Spencer Whitehead, Alexander~C. Berg, Wan{-}Yen Lo, Piotr Doll{\'{a}}r, and Ross~B. Girshick.
\newblock Segment anything.
\newblock In {\em {IEEE/CVF} International Conference on Computer Vision, {ICCV} 2023, Paris, France, October 1-6, 2023}, pages 3992--4003. {IEEE}, 2023.

\bibitem{li2021anytime_lns}
Jiaoyang Li, Zhe Chen, Daniel Harabor, Peter~J. Stuckey, and Sven Koenig.
\newblock Anytime multi-agent path finding via large neighborhood search.
\newblock In Zhi{-}Hua Zhou, editor, {\em Proceedings of the Thirtieth International Joint Conference on Artificial Intelligence, {IJCAI} 2021, Virtual Event / Montreal, Canada, 19-27 August 2021}, pages 4127--4135. ijcai.org, 2021.

\bibitem{li2022mapf-lns2}
Jiaoyang Li, Zhe Chen, Daniel Harabor, Peter~J. Stuckey, and Sven Koenig.
\newblock Mapf-lns2: Fast repairing for multi-agent path finding via large neighborhood search.
\newblock {\em Proceedings of the AAAI Conference on Artificial Intelligence}, 36(9):10256--10265, Jun. 2022.

\bibitem{srli2021}
Jiaoyang Li, Daniel Harabor, Peter~J. Stuckey, and Sven Koenig.
\newblock Pairwise symmetry reasoning for multi-agent path finding search.
\newblock {\em CoRR}, abs/2103.07116, 2021.

\bibitem{li2021eecbs}
Jiaoyang Li, Wheeler Ruml, and Sven Koenig.
\newblock Eecbs: A bounded-suboptimal search for multi-agent path finding.
\newblock In {\em Proceedings of the AAAI Conference on Artificial Intelligence (AAAI)}, pages 12353--12362, 2021.

\bibitem{li2020gnn}
Qingbiao Li, Fernando Gama, Alejandro Ribeiro, and Amanda Prorok.
\newblock Graph neural networks for decentralized multi-robot path planning, 2020.

\bibitem{li2021magat}
Qingbiao Li, Weizhe Lin, Zhe Liu, and Amanda Prorok.
\newblock Message-aware graph attention networks for large-scale multi-robot path planning, 2021.

\bibitem{li2022pico}
Wenhao Li, Hongjun Chen, Bo~Jin, Wenzhe Tan, Hongyuan Zha, and Xiangfeng Wang.
\newblock Multi-agent path finding with prioritized communication learning.
\newblock In {\em 2022 International Conference on Robotics and Automation (ICRA)}, pages 10695--10701, 2022.

\bibitem{lin2023sacha}
Qiushi Lin and Hang Ma.
\newblock {SACHA:} soft actor-critic with heuristic-based attention for partially observable multi-agent path finding.
\newblock {\em {IEEE} Robotics Autom. Lett.}, 8(8):5100--5107, 2023.

\bibitem{lin2024coco_dataset}
Tsung{-}Yi Lin, Michael Maire, Serge~J. Belongie, Lubomir~D. Bourdev, Ross~B. Girshick, James Hays, Pietro Perona, Deva Ramanan, Piotr Doll{\'{a}}r, and C.~Lawrence Zitnick.
\newblock Microsoft {COCO:} common objects in context.
\newblock {\em CoRR}, abs/1405.0312, 2014.

\bibitem{ma2021dhc}
Ziyuan Ma, Yudong Luo, and Hang Ma.
\newblock Distributed heuristic multi-agent path finding with communication.
\newblock In {\em 2021 IEEE International Conference on Robotics and Automation (ICRA)}, pages 8699--8705, 2021.

\bibitem{ma2022dcc}
Ziyuan Ma, Yudong Luo, and Jia Pan.
\newblock Learning selective communication for multi-agent path finding.
\newblock {\em IEEE Robotics and Automation Letters}, 7(2):1455--1462, 2022.

\bibitem{okumura2023lacam2}
Keisuke Okumura.
\newblock Improving lacam for scalable eventually optimal multi-agent pathfinding.
\newblock In {\em Proceedings of the Thirty-First International Joint Conference on Artificial Intelligence (IJCAI)}, 2023.

\bibitem{okumura2023lacam}
Keisuke Okumura.
\newblock Lacam: Search-based algorithm for quick multi-agent pathfinding.
\newblock In {\em Proceedings of AAAI Conference on Artificial Intelligence (AAAI)}, 2023.

\bibitem{okumura2024lacam3}
Keisuke Okumura.
\newblock Engineering lacam$^\ast$: Towards real-time, large-scale, and near-optimal multi-agent pathfinding.
\newblock In {\em Proceedings of International Conference on Autonomous Agents and Multiagent Systems (AAMAS)}, 2024.

\bibitem{okumara2022pibt_journal}
Keisuke Okumura, Manao Machida, Xavier Défago, and Yasumasa Tamura.
\newblock Priority inheritance with backtracking for iterative multi-agent path finding.
\newblock {\em Artificial Intelligence}, 310:103752, 2022.

\bibitem{sipp2011}
Mike Phillips and Maxim Likhachev.
\newblock Sipp: Safe interval path planning for dynamic environments.
\newblock In {\em 2011 IEEE International Conference on Robotics and Automation}, pages 5628--5635, 2011.

\bibitem{radford2021clip_model}
Alec Radford, Jong~Wook Kim, Chris Hallacy, Aditya Ramesh, Gabriel Goh, Sandhini Agarwal, Girish Sastry, Amanda Askell, Pamela Mishkin, Jack Clark, Gretchen Krueger, and Ilya Sutskever.
\newblock Learning transferable visual models from natural language supervision.
\newblock In Marina Meila and Tong Zhang, editors, {\em Proceedings of the 38th International Conference on Machine Learning, {ICML} 2021, 18-24 July 2021, Virtual Event}, volume 139 of {\em Proceedings of Machine Learning Research}, pages 8748--8763. {PMLR}, 2021.

\bibitem{radford2019gpt2_webtext}
Alec Radford, Jeffrey Wu, Rewon Child, David Luan, Dario Amodei, Ilya Sutskever, et~al.
\newblock Language models are unsupervised multitask learners.

\bibitem{raffel2019colossalcrawledcorpus}
Colin Raffel, Noam Shazeer, Adam Roberts, Katherine Lee, Sharan Narang, Michael Matena, Yanqi Zhou, Wei Li, and Peter~J. Liu.
\newblock Exploring the limits of transfer learning with a unified text-to-text transformer.
\newblock {\em CoRR}, abs/1910.10683, 2019.

\bibitem{ross2011dagger}
St{\'{e}}phane Ross, Geoffrey~J. Gordon, and Drew Bagnell.
\newblock A reduction of imitation learning and structured prediction to no-regret online learning.
\newblock In Geoffrey~J. Gordon, David~B. Dunson, and Miroslav Dud{\'{\i}}k, editors, {\em Proceedings of the Fourteenth International Conference on Artificial Intelligence and Statistics, {AISTATS} 2011, Fort Lauderdale, USA, April 11-13, 2011}, volume~15 of {\em {JMLR} Proceedings}, pages 627--635. JMLR.org, 2011.

\bibitem{sartoretti2019primal}
Guillaume Sartoretti, Justin Kerr, Yunfei Shi, Glenn Wagner, TK~Satish Kumar, Sven Koenig, and Howie Choset.
\newblock Primal: Pathfinding via reinforcement and imitation multi-agent learning.
\newblock {\em IEEE Robotics and Automation Letters}, 4(3):2378--2385, 2019.

\bibitem{sharon2015cbs}
Guni Sharon, Roni Stern, Ariel Felner, and Nathan~R Sturtevant.
\newblock Conflict-based search for optimal multi-agent pathfinding.
\newblock {\em Artificial Intelligence}, 219:40--66, 2015.

\bibitem{skrynnik2024learn-to-follow}
Alexey Skrynnik, Anton Andreychuk, Maria Nesterova, Konstantin Yakovlev, and Aleksandr Panov.
\newblock Learn to follow: Decentralized lifelong multi-agent pathfinding via planning and learning.
\newblock In {\em Proceedings of the AAAI Conference on Artificial Intelligence}, volume~38, pages 17541--17549, 2024.

\bibitem{stern2019mapfbenchmark}
Roni Stern, Nathan~R. Sturtevant, Ariel Felner, Sven Koenig, Hang Ma, Thayne~T. Walker, Jiaoyang Li, Dor Atzmon, Liron Cohen, T.~K.~Satish Kumar, Eli Boyarski, and Roman Bartak.
\newblock Multi-agent pathfinding: Definitions, variants, and benchmarks.
\newblock {\em Symposium on Combinatorial Search (SoCS)}, pages 151--158, 2019.

\bibitem{tang2024eph}
Huijie Tang, Federico Berto, and Jinkyoo Park.
\newblock Ensembling prioritized hybrid policies for multi-agent pathfinding.
\newblock In {\em 2024 IEEE/RSJ International Conference on Intelligent Robots and Systems (IROS)}. IEEE, 2024.
\newblock \url{https://github.com/ai4co/eph-mapf}.

\bibitem{effectiveCBS}
Rishi Veerapaneni, Tushar Kusnur, and Maxim Likhachev.
\newblock Effective integration of weighted cost-to-go and conflict heuristic within suboptimal {CBS}.
\newblock In {\em Thirty-Seventh {AAAI} Conference on Artificial Intelligence, {AAAI} 2023, Washington, DC, USA, February 7-14, 2023}, pages 11691--11698. {AAAI} Press, 2023.

\bibitem{veerapaneni2024improving_mapf_policies_with_search}
Rishi Veerapaneni, Qian Wang, Kevin Ren, Arthur Jakobsson, Jiaoyang Li, and Maxim Likhachev.
\newblock Improving learnt local mapf policies with heuristic search.
\newblock {\em International Conference on Automated Planning and Scheduling}, 34(1):597--606, 2024.

\bibitem{wang2023scrimp}
Yutong Wang, Bairan Xiang, Shinan Huang, and Guillaume Sartoretti.
\newblock Scrimp: Scalable communication for reinforcement- and imitation-learning-based multi-agent pathfinding, 2023.

\bibitem{yan2024neural-lns}
Zhongxia Yan and Cathy Wu.
\newblock Neural neighborhood search for multi-agent path finding.
\newblock In {\em International Conference on Learning Representations}, 2024.

\bibitem{yu2023learnConflictHeuristic}
Chenning Yu, Qingbiao Li, Sicun Gao, and Amanda Prorok.
\newblock Accelerating multi-agent planning using graph transformers with bounded suboptimality.
\newblock In {\em 2023 IEEE International Conference on Robotics and Automation (ICRA)}, pages 3432--3439, 2023.

\bibitem{zhang2024ggo}
Yulun Zhang, He~Jiang, Varun Bhatt, Stefanos Nikolaidis, and Jiaoyang Li.
\newblock Guidance graph optimization for lifelong multi-agent path finding.
\newblock In {\em Proceedings of the International Joint Conference on Artificial Intelligence (IJCAI)}, pages 311--320, 2024.

\bibitem{zhu2015bookcorpus}
Yukun Zhu, Ryan Kiros, Rich Zemel, Ruslan Salakhutdinov, Raquel Urtasun, Antonio Torralba, and Sanja Fidler.
\newblock Aligning books and movies: Towards story-like visual explanations by watching movies and reading books.
\newblock In {\em The IEEE International Conference on Computer Vision (ICCV)}, December 2015.

\end{thebibliography}

\clearpage
\setcounter{figure}{0}
\renewcommand{\thefigure}{A\arabic{figure}}
\setcounter{table}{0}
\renewcommand{\thetable}{A\arabic{table}}
\setcounter{section}{0}
\renewcommand{\thesection}{A\arabic{section}}

\section*{Appendix}

\section{Quick Summary}
\subsubsection{Recommended Background Reading} 
Readers unfamiliar with Multi-Agent Path Finding or variants are recommended to read \cite{stern2019mapfbenchmark}. Readers unfamiliar with Conflict Based Search or suboptimal variants should read ECBS \cite{barer2014suboptimalecbs}. Readers unfamiliar with machine learning-based approaches for MAPF should read PRIMAL \cite{sartoretti2019primal} and PRIMAL2 \cite{damani2021primal2} for reinforcement learning-based approaches, and GNN \cite{li2020gnn} and MAGAT \cite{li2021magat} for imitation learning approaches. Readers interested in collision shielding should read \cite{veerapaneni2024improving_mapf_policies_with_search}.

\subsubsection{Motivation with respect to prior work}
The majority of prior papers that have attempted to learn decentralized MAPF policies use reinforcement learning. However under current set-ups, they typically train on relatively few agents (e.g. 16 or 100), and on a limited variety of maps. Prior supervised methods used relatively weak heuristic search solvers. 

Our initial goal was to demonstrate the effective use of simple and scalable imitation learning to learn a state-of-the-art decentralized MAPF policy that works across a variety of maps. Our findings however show a different story.

\subsection{Main Takeaways} 
1. ``Work Smarter not Harder" using CS-PIBT: We trained a GNN MAPF policy on a large dataset (over 700k examples, which contain hundreds of agents). However, when using naive collision shielding which replaces colliding predicted actions with wait actions, performance was not particularly impressive. When running the same model with CS-PIBT, we could get unquestionably ``state-of-the-art" ML MAPF performance. Additionally, we saw that large scale data was not necessary and that we could get the same performance with 1/100th the data with training time taking minutes. \textbf{All future ML MAPF policies (with local communication) should use a smart collision shield (i.e. CS-PIBT)}. Local communication is needed for CS-PIBT.

2. Using a smart collision shield (i.e. CS-PIBT) fundamentally changes the learning problem: The difference in performance between using a naive collision shield and CS-PIBT reveals that a key bottleneck in existing ML MAPF policies is resolving 1-step collisions. Using CS-PIBT eliminates this issue. We also qualitatively noticed that our model with CS-PIBT failed in instances where longer-horizon planning is required (e.g. target conflicts). Thus, since all applicable future ML MAPF policies should use a smart collision shield, this means that \textbf{future ML MAPF policies can focus on longer-horizon reasoning}.

3. Use appropriate baselines: Existing decentralized ML MAPF policies have failed to compare against PIBT which is a decentralized heuristic search baseline. Related, since we are advocating using CS-PIBT, a trivial solution for the ML MAPF policy is to learn a greedy policy that minimizes the heuristic. Thus, when using CS-PIBT, \textbf{future works should always compare against CS-PIBT with a greedy heuristic (i.e. PIBT)}.

4. Simple and Scalable Imitation Learning is feasible: We can leverage strong centralized heuristic search solvers to collect a large dataset across a diverse set of maps. The runtime of collecting a dataset is directly proportional to the number of parallel jobs that can be run, with 64 cores collecting this large dataset took 90 minutes to collect. Note we did not parallelize converting the raw paths into a pytorch dataset which took approximately 6 hours accordingly. 
Prior works have largely used RL that can only train on with relatively low agent density / congestion. 

We see that without CS-PIBT (i.e. using CS-Naive), increasing the amount of training data leads to improvements in model performance. We additionally notice  that with CS-PIBT, although training the same model with more data did not noticeably change the success rate, it did improve the solution cost per agent (e.g. Paris\_1\_256 in Fig \ref{fig:main-results}). Future works will likely find it useful to utilize large scale imitation learning for learning long-horizon behaviors.

\section{Additional Information on Experiments}

\subsection{Training Data Information}
We split the Moving AI Benchmark into a training and test set such that at least one of each map type (city, empty, maze, random, warehouse, game) would be in the test set. Thus, our test set consisted of the following maps: Paris\_1\_256, empty\_48\_48, maze\_128\_128\_2, random\_64\_64\_10, random\_32\_32\_10, warehouse\_10\_20\_10\_2\_1, den312d, den520d. All other maps were used for training, plus the addition of a few random maps as mentioned in Section \ref{sec:dataset}.

We used and modified EECBS publicly available code to collect our training data. Instances where EECBS did not find solutions (i.e. timed out) were not included in the training data. Since we assumed timesteps were independent (a limitation of our model discussed in Section \ref{sec:limitations}), each timestep on the MAPF solution corresponds to its own training example (as the label for each agent is the next action they took). 

One minor detail is that a large amount of labels are ``wait" actions (about 50\% of our dataset). This occurs as a MAPF solution for many agents usually has most agents waiting at their goal mid-way and a few agents left navigating at the end. We did not find it necessary to throw away wait actions or down-weight wait labels.

We tried DAgger \cite{ross2011dagger} to obtain more data given our current SSIL with CS-PIBT performance. Given a run of SSIL with CS-PIBT (regardless of if it succeeds or fails), we sample 10 timesteps and feed in each timestep as the ``start" locations to EECBS to get more MAPF solution paths. We can then add this to the dataset. We experimented with various dataset balancing mechanisms, e.g. sampling more from failure instances or later in the solution, but did not find any to significantly boost performance.

\subsection{Model Architecture}
We employ a SageConv \cite{hamilton2017sagegraphconv} graph neural network. Nodes in the graph correspond to agents with edges denoting local communication between agents within the field of view.

We first recap the inputs. We do not use edge features. The node features are:
\begin{enumerate}
    \item 3 $D*D$ images centered at the current agent: one binary obstacle map, one normalized Backward Dijkstra (BD) heuristic, one binary agent occupancy.
    \item A size 5 binary vector corresponding to the 5 actions with a 1 if that action minimizes the BD heuristic.
\end{enumerate}

Each graph convolutional layer requires a round of message passing between agents. This first requires each agent to convert its features into two latent vectors, one for itself while another that it sends to neighboring agents.

We first encode the $(3,D=9,D=9)$ image with a $(3,3,3)$ CNN with stride 1 and no padding. We then flatten the embedding and concatenate the size 5 binary vector.
We then apply one linear layer to get a node embedding of size 128. We similarly apply a different linear layer to obtain a message vector of size 128 that is passed to neighbors.

We then apply 3 SageConv graph convolutional layers with intermediate layer norms. We finally apply a linear layer of size 128, dropout of 0.25, and then output the action probability distribution of size 5. We used intermediate LeakyRelu activations and Layer Norms after the SageConv layers.

We found that the most critical component was using the SageConv layers as opposed to others we tried. Changing the number of layers, sizes, or activations did not significantly change our success rate with CS-PIBT.

\subsection{Full Benchmark Results}
\begin{figure*}[t!]
    \centering
    \includegraphics[width=0.99\linewidth]{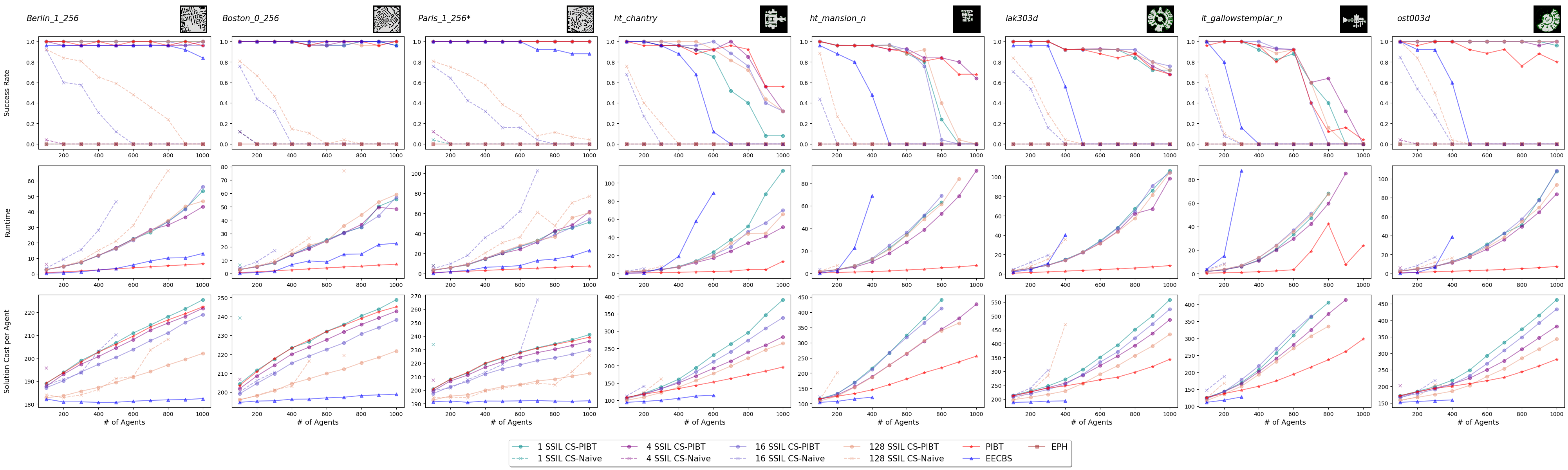}
    \vspace{-0.5em}
    \caption{Success, solution cost, and runtime for city and game maps ($^*$ denotes test maps).}
    \label{fig:city-maps-appendix}
    \vspace{-1em}
\end{figure*}

\begin{figure*}[t!]
    \centering
    \includegraphics[width=0.99\linewidth]{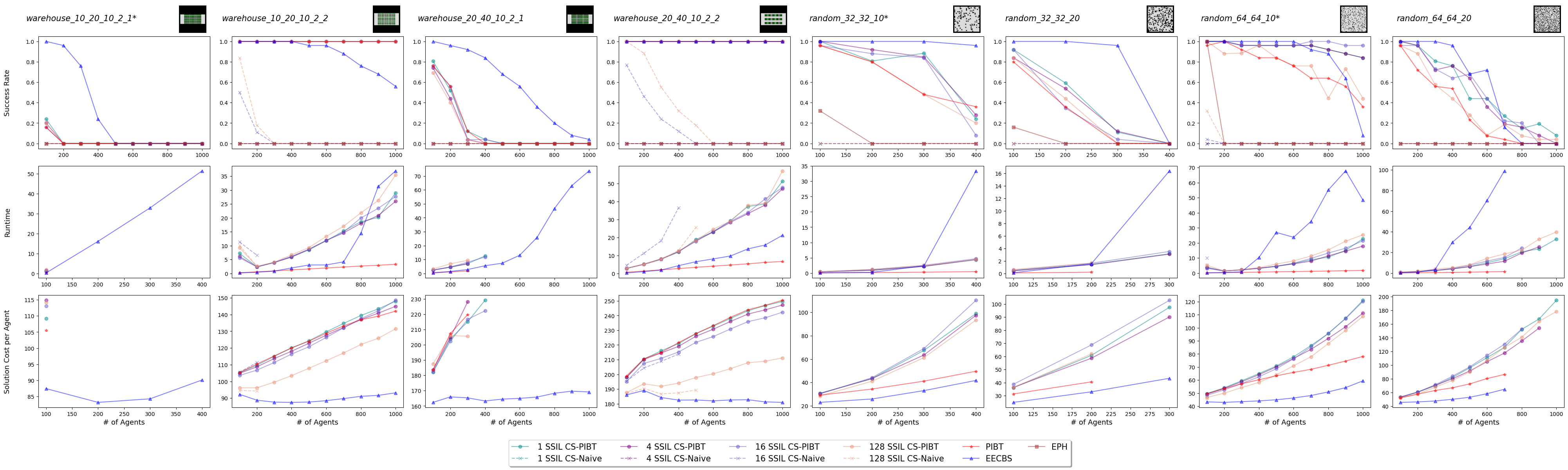}
    \vspace{-0.5em}
    \caption{Success, solution cost, and runtime for warehouse and random maps ($^*$ denotes test maps).}
    \label{fig:warehouse-maps-appendix}
    \vspace{-1em}
\end{figure*}

\begin{figure*}[t!]
    \centering
    \includegraphics[width=0.99\linewidth]{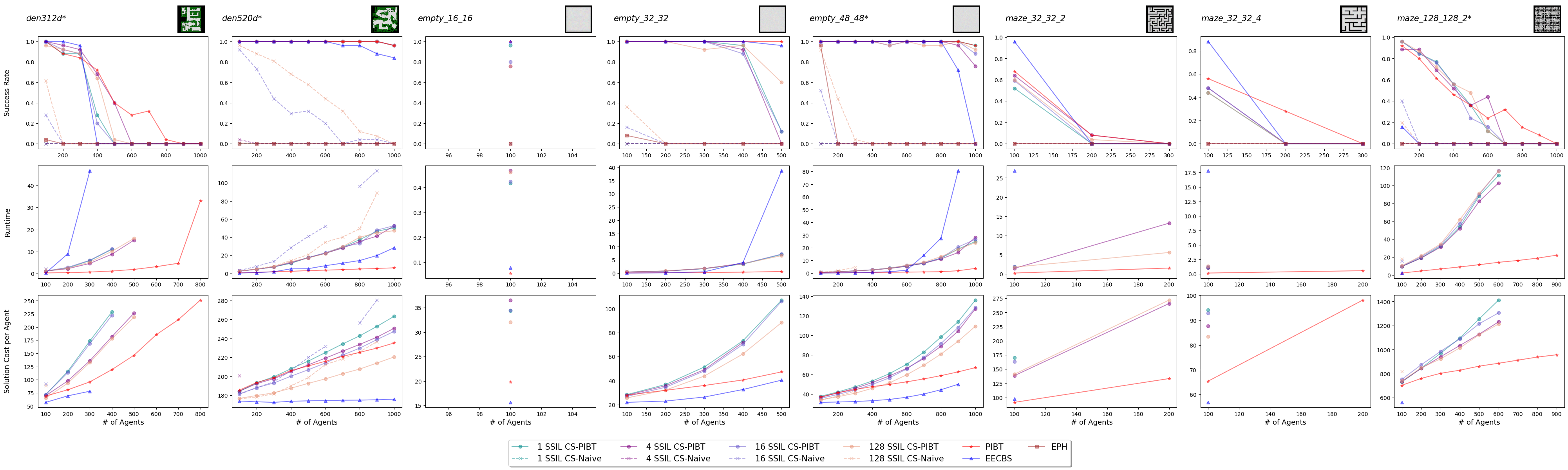}
    \vspace{-0.5em}
    \caption{Success, solution cost, and runtime for maze, den, and empty maps ($^*$ denotes test maps).}
    \label{fig:maze-maps-appendix}
    \vspace{-1em}
\end{figure*}

\begin{figure*}[t!]
    \centering
    \includegraphics[width=0.35\linewidth]{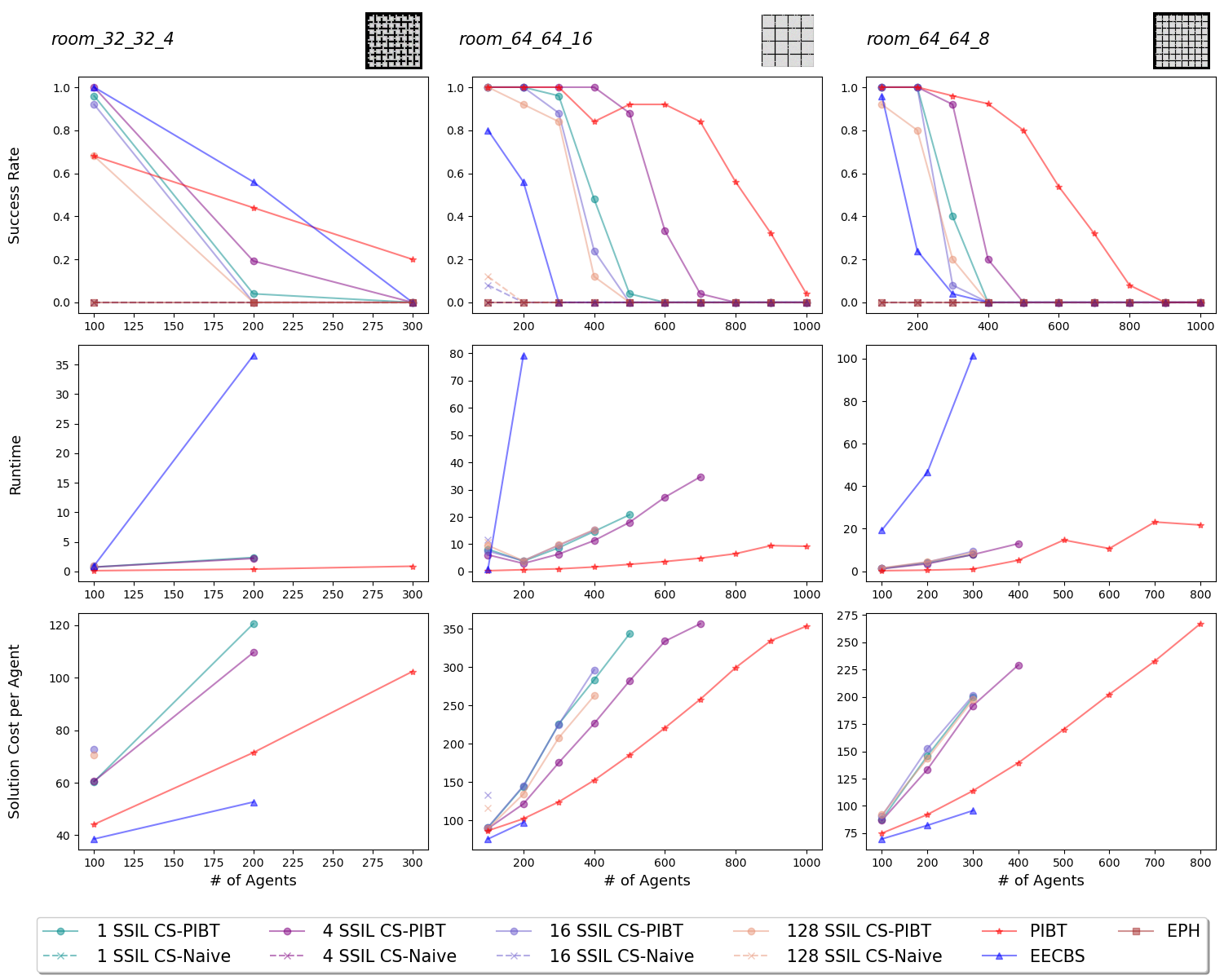}
    \vspace{-0.5em}
    \caption{Success, solution cost, and runtime for room maps.}
    \label{fig:room-maps-appendix}
    \vspace{-1em}
\end{figure*}

Figure \ref{fig:city-maps-appendix}, \ref{fig:warehouse-maps-appendix}, \ref{fig:maze-maps-appendix}, \ref{fig:room-maps-appendix} show the full results across the entire Moving AI benchmark grouped by map types (except for the four we removed as mentioned earlier). Maps with $^*$ denotes test maps that we did not see during training. 
We observe identical patterns as shown in \ref{fig:main-results} and therefore do not repeat the analysis described in Section \ref{sec:results}.

\subsubsection{Runtime}
We implemented SSIL and CS-PIBT/PIBT in Python. We note that the runtime seen in the graphs is the cumulative runtime across all timesteps. Thus, even though the cumulative runtime may be in seconds, this is the sum across hundreds of timesteps/actions and the runtime per action is substantially smaller. SSIL with CS-PIBT is quite fast, e.g. running 1000 agents in den312d taking about $0.06$ seconds per collision-free actions (for all 1000 agents). The majority of this runtime (over 60\%) is creating the input features (e.g. the FoV images) for the model rather than the model inference itself. CS-PIBT can take a non-negligible amount depending on the amount of inter-agent conflicts (e.g. about 20\% of the per-iteration runtime). Thus model inference time only takes around 20\% of total time. We do not use a GPU during execution as the data loading time slowed down the process.

We note that our model runs on significantly more agents/congestion than prior work. Prior works have omitted runtime information or have different scaling performance (e.g. theirs could be faster with 50 agents but significantly slower with 200) so it is hard to compare runtimes.

\end{document}